\documentclass[fleqn,10pt]{wlscirep}
\usepackage[utf8]{inputenc}
\usepackage[T1]{fontenc}

\usepackage{color}
\usepackage{graphicx}
\usepackage{longtable}
\usepackage{amsmath}
\usepackage{amsthm}
\usepackage{amssymb}
\usepackage{lipsum}
\usepackage{latexsym}
\usepackage{bm}

\definecolor{lightblue}{rgb}{.0, .0, .8}

\newcommand{\lla}{\left\langle}
\newcommand{\rra}{\right\rangle}
\newcommand{\psum}{\sideset{}{'}\sum}

\newcommand{\dV}{\Delta V}
\newcommand{\vGamma}{\varGamma}
\newcommand{\REV}[1]{\textcolor{black}{#1}}

\newcommand{\ocite}[1]{\hspace{-1 ex} \nocite{#1}\citenum{#1}}

\title{Local stress and pressure in an inhomogeneous system of spherical active Brownian particles}

\author{Shibananda Das}
\author{Gerhard Gompper}
\author{Roland G. Winkler}
\affil{Theoretical Soft Matter and Biophysics, Institute of Complex Systems and Institute for Advanced
Simulation, Forschungszentrum J\"{u}lich, 52425 J\"{u}lich, Germany}
\affil[*]{r.winkler@fz-juelich.de, g.gompper@fz-juelich.de}

\begin{abstract}
The stress of a fluid on a confining wall is given by the mechanical wall forces, independent of the nature of the fluid being passive or active. At thermal equilibrium, an equation of state exists and stress is likewise obtained from intrinsic bulk properties; even more, stress can be calculated locally. Comparable local descriptions for active systems require a particular consideration of active forces. Here, we derive expressions for the stress exerted on a local volume of a systems of spherical active Brownian particles (ABPs). Using the virial theorem, we obtain two identical stress expressions,  a stress due to momentum flux across a hypothetical plane, and a bulk stress inside of the local volume. In the first case, we obtain an active contribution to momentum transport in analogy to momentum transport in an underdamped passive system, and we introduce an active momentum. In the second case, a generally valid expression for the swim stress is derived. By simulations, we demonstrate that the local bulk stress is identical to the wall stress of a confined system for both, non-interacting ABPs as well as ABPs with excluded-volume interactions. This underlines the existence of an equation of state for a system of spherical ABPs. Most importantly, our calculations demonstrated that active stress is not a wall (boundary) effect, but is caused by momentum transport. We demonstrate that the derived stress expression permits the calculation of the local stress in inhomogeneous systems of ABPs.
\end{abstract}

\begin{document}

\flushbottom
\maketitle
%
%
\thispagestyle{empty}

\section*{Introduction}

The agents of active matter perpetually convert internal energy or energy from the environment into systematic translational motion  and are, thus, usually far from thermal equilibrium \cite{rama:10,cate:11,marc:13,elge:15,bech:16}. Their motility gives rise to remarkable phenomena such as collective motion, activity-induced phase transitions \cite{fily:12,theu:12,butt:13,redn:13,bial:12,witt:14,cate:15,wyso:14,sten:14,wyso:16,marc:16.1,bech:16,thee:18}, or wall accumulation \cite{berk:08,lee:13,elge:13.1,fily:14}. Activity is associated with the generation of active stresses, which are responsible for the persistent nonequilibrium state of a system \cite{taka:14,yang:14.2,solo:15}. Such stresses are particularly interesting, since there is no equilibrium counterpart and thus, they usually cannot be described by equilibrium thermodynamics. An expression for the active pressure in systems of active Brownian particles (ABPs), based on a generalization of the virial including the active force, has been introduced recently \cite{taka:14,yang:14.2}.
Subsequently, expressions for the active pressure have been derived in various ways for both, the overdamped dynamics of ABPs \cite{wink:15,fala:16,spec:16,solo:15.1} as well as in presence of inertia \cite{joye:16,taka:17}.

Mechanical stress/pressure as force per area  can be defined even far from equilibrium. The fundamental question is then, how this mechanical pressure is related to nonequilibrium thermodynamics. As a step toward such a relation, the existence of an equation of state for systems of ABPs has been addressed \cite{solo:15,solo:15.1,gino:15}. Various simulation and theoretical studies confirm that indeed such a relation, including the active pressure, exists for spherical ABPs \cite{yang:14.2,solo:15,taka:14,gino:15,bert:15,wink:15,niko:16}. However, for (anisotropic) active particles which experience a torque, pressure is in general not a state function  anymore \cite{solo:15.1,juno:17}.

A pressure equation of state implies equivalence of wall and local pressure at any point inside the homogeneous relevant volume. Exactly this identity has been questioned in Ref.~\ocite{spec:16}. Typically, confined or periodic systems are considered and relations between wall forces, hence the wall pressure, and virial, including the active-force virial, are determined \cite{yang:14.2,solo:15.1,taka:14,wink:15}, but no local (bulk) pressure is calculated. Aside from Ref.~\ocite{spec:16}, local pressure and the existence of an equation of state has been addressed in Ref.~\ocite{fily:18}.

In this article, we derive an instantaneous expression for the local stress (pressure) in a system of spherical active Brownian particles using the virial approach. Thereby, we follow the strategy of Ref.~\ocite{lion:12} for passive systems. A volume $\dV$ within a larger volume $V$ is considered and an expression is derived for the pressure inside $\dV$. Considering the equations of motions of the ABPs inside the volume $\dV$, including inertia, we obtain the well-known virial expressions for the pressure of a passive system. In addition, however, we find  activity contributions. On the one hand, we obtain a virial expression for the active stress (swim stress) of particles inside $\dV$, which differs from the swim stress discussed so far, and, on the other hand, we find an original momentum-flux contribution of APBs to stress, which we denote as active momentum. The notation rests upon the similarity to the classical linear momentum of a passive particle. It is precisely this term, which distinguishes our studies from those of Ref.~\ocite{spec:16}. In general, we agree with the arguments presented in Ref.~\ocite{spec:16}, however, the momentum flux across a plane in the system, similar to the  momentum flux by inertia, yields an extra contribution to the local stress which not considered in Ref.~\ocite{spec:16}. \REV{In fact, in Ref.~\ocite{fily:18} an ``active impulse, defined as the mean momentum a particle will receive on average from the substrate in the future'', is introduce, resembling our instantaneous active momentum, but the meaning is different.} By computer simulations of non-interacting  ABPs and ABPs interacting by a Lennard-Jones potential, we demonstrate that the derived expressions for the wall and bulk stress are identical. This highlights the existence of a pressure equation state for spherical ABP systems \cite{taka:14,solo:15,fily:18}. Moreover, it opens up the possibility to calculate stresses locally, even in inhomogeneous systems, which we confirm by calculating the local stress in a phase separated system.

\section*{Stress in active systems} \label{sec:abp_model}
\subsection*{Model}
We consider a system of $N$  ABPs of diameter $\sigma$ in a three-dimensional volume $V$. Their equations of motion, including the inertia contribution, are given by
\begin{align} \label{eq:abp_trans}
m \ddot{\bm r_i} + \gamma \dot{\bm r_i} (t)= \gamma v_{0}{\bm e}_i (t)+ {\bm F}_i(t)+ {\bm \vGamma}_i(t) ,
\end{align}
where ${\bm r}_i$, $\dot {\bm r_i}$, and ${\bm F}_i$ ($i=1,\ldots,N$) denote the particle positions,  velocities, and forces, respectively,  $m$ is the mass, $\gamma$ the translational friction coefficient, and ${\bm \vGamma}_i$ are Gaussian white-noise random forces, with the moments
\begin{align}
\lla {\bm \vGamma}_i (t)\rra  = 0,   \hspace*{3mm} \lla {\vGamma}_{i \alpha}(t)  {\vGamma}_{j \beta} (t') \rra = 2 \gamma k_B T \delta_{\alpha \beta} \delta_{ij} \delta(t-t') .
\end{align}
Here, $T$ denotes the temperature, $k_B$ the Boltzmann constant, and $\alpha,  \beta \in \{x,y,z\}$ the Cartesian coordinate directions.
A particle $i$ is propelled  with constant velocity $v_{0}$ along its orientation vector ${\bm e}_i$. The latter changes in a diffusive manner according to
\REV{\begin{align} \label{eq:orient}
\dot {\bm e_i}(t)= {\bm e}_i(t) \times {\bm \eta}_i(t) - 2 D_R \bm e_i
\end{align}
within the Ito interpretation of the multiplicative noise process \cite{raib:04};
$\bm \eta_i$ is a Gaussian and Markovian stochastic process, with the moments}
\begin{align}
\langle {\eta_{i \alpha}} (t) \rangle =  0  , \hspace*{3mm} \langle {\eta}_{i \alpha}(t)  {\eta}_{j \beta}(t')\rangle =   \gamma_R \delta_{\alpha \beta} \delta_{ij} \delta(t-t') .
\end{align}
The damping factor $\gamma_R$ is related to the rotational diffusion coefficient, $D_R$, via $\gamma_R = 2 D_R$. Alternatively, the velocity equation of motion of an active Ornstein-Uhlenbeck particle can be considered \cite{das:18.1,fodo:16}, where aside from the orientation also the magnitude of the propulsion velocity is changing. In any case, the correlation function for the active velocity $\bm v_i^a=v_0\bm e_i$ is \cite{das:18.1,wink:15}
\begin{align}
\lla \bm v_i^a(t) \cdot \bm v_i^a(0) \rra = v_0^2\lla \bm e_i (t) \cdot \bm e_i(0) \rra=  v_0^2e^{-\gamma_R t}  .
\end{align}
The forces $\bm F_i$ include contributions from (short-range) interactions with confining walls, $\bm F_i^w$, as well as possible pair-wise ABP interactions, $\bm F_{ij}$, such that (the prime at the sum indicates that the index $j = i$ is excluded)
\begin{align} \label{eq:corr_e}
\bm F_i = \bm F_i^w + \psum_{j=1}^N \bm F_{ij} .
\end{align}

\subsection*{\REV{Global} stress in confined system}

We examine now ABPs confined in a rectangle (2D) or cuboid (3D) with walls located at $\pm L_{\alpha}/2$. The virial expression for the stress follows by multiplying the respective individual Cartesian equations of Eqs.~(\ref{eq:abp_trans}) by $r_{i \alpha}$, which yields \cite{wink:93,wink:15}
\begin{align} \label{eq:virial_surf}
\frac{d}{dt} \left[m \lla \dot{r}_{i \alpha} r_{i \alpha} \rra + \frac{\gamma}{2}  \lla r_{i \alpha}^2 \rra \right] =   \ m \lla \dot{r}_{i \alpha}^2 \rra +\gamma  \lla v^a_{i \alpha} r_{i \alpha} \rra + \lla F^w_{i \alpha} r_{i \alpha} \rra +
\psum_{j=1}^N \lla F_{ij \alpha}r_{i \alpha}  \rra ,
\end{align}
where $\lla \ldots \rra$ denotes either a time or an ensemble average. Due to confinement, the left-hand side of Eq.~(\ref{eq:virial_surf}) vanishes.  With the definition of an external stress $\sigma_{\alpha \alpha}^e$ as average total force exerted on a wall,
\begin{align} \label{eq:stress_ext}
V \sigma_{\alpha \alpha}^e =  \sum_{i=1}^N \lla F_{i \alpha}^{w} S_{i \alpha} \rra,
\end{align}
where  $S_{i \alpha} = \pm L_\alpha/2$  denotes the location of the wall with which particle $i$ interacts,
Eq.~(\ref{eq:virial_surf}) yields the equivalent internal stress
\begin{align}  \label{eq:stress_int}
V\sigma_{\alpha \alpha}^i =  - \sum_{i=1}^N m \lla \dot{r}_{i \alpha}^2 \rra - \sum_{i=1}^N \gamma  \lla v^a_{i \alpha} r_{i \alpha} \rra   - \sum_{i=1}^N \lla F^w_{i \alpha}(r_{i \alpha}-S_{i\alpha}) \rra - \frac{1}{2} \sum_{i=1}^N \psum_{j=1}^N \lla F_{ij \alpha}r_{ij \alpha}  \rra ,
\end{align}
with $\bm r_{ij} = \bm r_i - \bm r_j$, \REV{as sum over all particles of the confined systems. Hence, we denote this expression as global (wall) stress.}  Note that by definition  $\sigma_{\alpha \alpha}^e =\sigma_{\alpha \alpha}^i$. The contribution with the active force is denoted as swim stress \cite{taka:14,taka:14.1, wink:15}
\begin{align} \label{eq:swim_stress_trad}
V \sigma_{\alpha \alpha}^s = - \sum_{i=1}^N \gamma  \lla v^a_{i \alpha} r_{i \alpha} \rra .
\end{align}
The pressure  itself follows as $p^{i/e} = -  (\sum_{\alpha} \sigma_{\alpha \alpha}^{i/e})/3$.
Aside from the wall term with the forces $\bm F_i^w$,  the stress tensor (\ref{eq:stress_int}) is identical with that derived in Ref.~\ocite{wink:15}. The wall term captures the finite range of the wall force and vanishes in the limit of vanishing wall interaction range.

\begin{figure}[t]
\centering
\includegraphics[width=0.9\columnwidth]{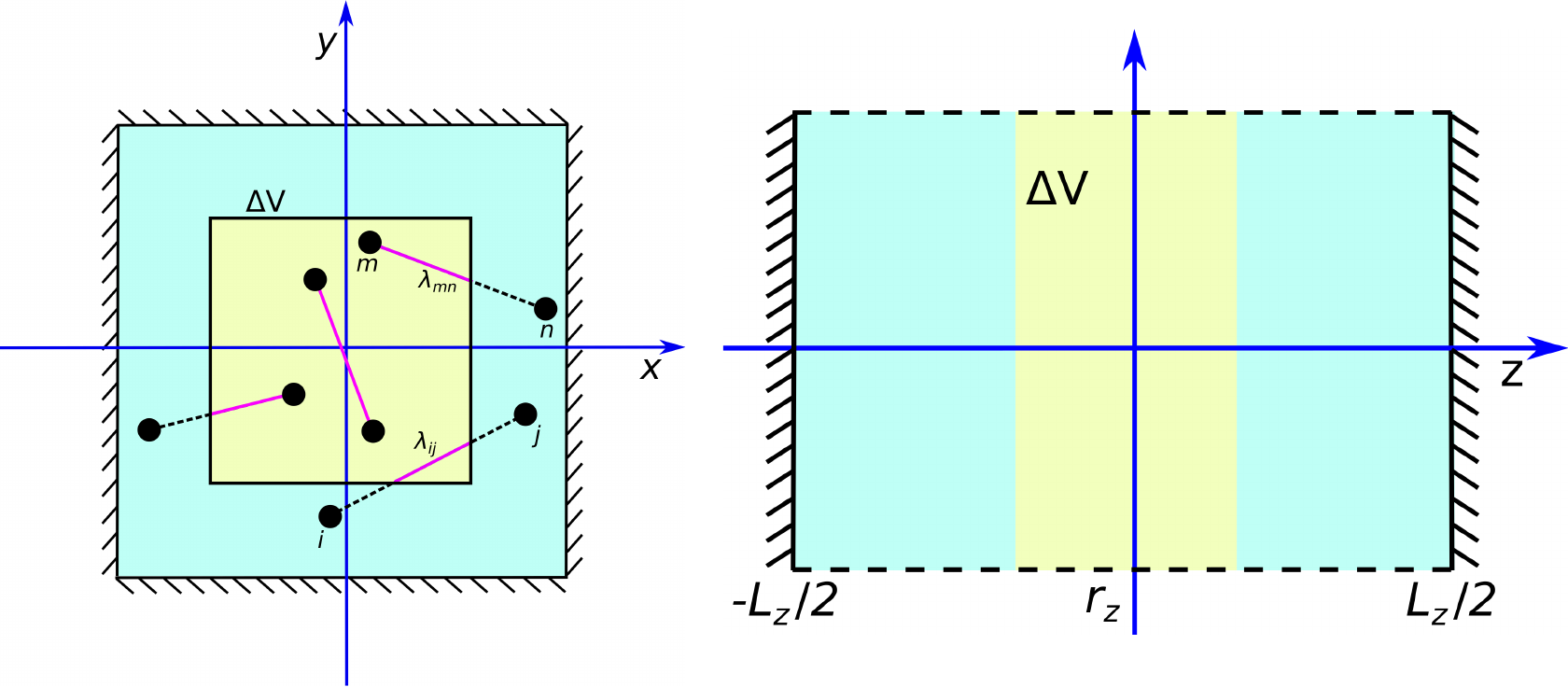}
\caption{(Left) Schematic of the local subvolume $\Delta V$ (yellow) embedded in the bulk of a system confined between walls (light blue). The fraction $\lambda_{ij}$ of the line joining particle $i$ and $j$, which lies inside $\dV$, is indicated by a solid line. (Right) Schematic of the subvolume $\Delta V$ for the calculation of local stress along the $z$-axis in the simulations. The walls are separated by $L_z$, and periodic boundaries are applied in the other two directions. The center position of $\dV$ along the $z$-axis is denoted by $r_z$.}
\label{fig:local_region}
\end{figure}

\subsection*{Local virial and local stress} \label{sec:local_pressure}

In order to determine the local stress at the position $\bm r$ within the volume $V$, we consider a \REV{cuboidal subvolume $\dV= \Delta L_{x} \times \Delta L_{y}  \times \Delta L_{z} $} centered around $\bm r$,  \REV{with the side lengths $\Delta L_{\alpha}$}  (cf. Fig.~\ref{fig:local_region}). ABPs inside of $\Delta V$ interact with each other as well as ABPs outside. Moreover, ABPs enter and leave $\dV$ in the course of time. The internal stress follows from the suitable virial expression of ABPs inside of $\dV$. To derive the virial expression, as before, we multiply the  equations~(\ref{eq:abp_trans})  and (\ref{eq:orient}) by $r_{i \alpha}$, an additional factor $\Lambda_{i \alpha}(\bm r)$, and sum over all particles.  The factor $\Lambda_{i \alpha}$ determines the volume $\dV$---$\Lambda_{i \alpha}(\bm r)$ is unity when particle $i$ is within $\dV$ in the coordinate direction $\alpha$ and zero otherwise \cite{lion:12}.  As a result, averaging yields
\begin{align} \label{eq:virial_cm}
\sum_{i=1}^N m \lla \dot{r}_i^2 \Lambda_{i} \rra +\sum_{i=1}^N m \lla \dot{r}_{i}  r_{i} \dot \Lambda_{i} \rra +  \sum_{i=1}^N \lla F_{i} r_{i} \Lambda_{i} \rra +  \sum_{i=1}^N \lla \Gamma_{i} r_{i} \Lambda_{i} \rra + \gamma \sum_{i=1}^N  \lla  v^a_{i} r_{i} \Lambda_{i} \rra =0 ,
\\ \label{eq:virial_e}
\sum_{i=1}^N  \lla  e_{i} \dot r_{i} \Lambda_{i} \rra +\sum_{i=1}^N  \lla  e_{i} r_{i} \dot \Lambda_{i} \rra - \gamma_R \sum_{i=1}^N  \lla  e_{i} r_{i} \Lambda_{i} \rra=0 ,
\end{align}
with the Ito interpretation of the colored noise process for $\bm e$ \cite{risk:89,wink:15}, and the relations
\begin{align}  \label{eq:der_acc}
\frac{d}{dt} \lla \dot r_i r_i \Lambda_i \rra & = \lla \ddot r_i r_i \Lambda_i \rra + \lla \dot r_i^2 \Lambda_i \rra + \lla \dot r_i r_i \dot \Lambda_i \rra,  \hspace*{3mm} \frac{d}{dt} \lla  e_i r_i \Lambda_i  \rra  = \lla \dot e_i r_i \Lambda_i \rra + \lla e \dot r_i \Lambda_i \rra +  \lla e_i r_i \dot \Lambda_i \rra .
\end{align}
Here, and in the following, we often suppress the index $\alpha$  for compactness of the expressions.
Due to confinement, the left-hand sides of Eqs.~(\ref{eq:der_acc})  vanish. Similarly, the averages $\lla \dot r_i r_i \Lambda_i \rra = \langle r_i^2 \dot \Lambda_i \rangle =0$. Inserting the third term of Eq.~(\ref{eq:virial_e}) into Eq.~(\ref{eq:virial_cm}), we obtain the virial expression
\begin{align} \label{eq:virial}
\sum_{i=1}^N m \lla \dot{r}_i^2 \Lambda_{i} \rra +\sum_{i=1}^N m \lla \dot{r}_{i}  r_{i} \dot \Lambda_{i} \rra +  \sum_{i=1}^N \lla F_{i} r_{i} \Lambda_{i} \rra +  \frac{\gamma}{\gamma_R} \left[\sum_{i=1}^N  \lla v^a_{i} \dot r_{i} \Lambda_{i} \rra +\sum_{i=1}^N  \lla  v^a_{i} r_{i} \dot \Lambda_{i} \rra \right] =0 .
\end{align}
Here, we neglect  terms with the averages $\langle \Gamma_{i} r_{i} \Lambda_{i} \rangle$; these averages vanish as long as inertia (cf. Eq.~(\ref{eq:abp_trans})) is taken into account.
For an overdamped dynamics, $m=0$ in Eq.~(\ref{eq:abp_trans}), $\langle \Gamma_{i} r_{i} \Lambda_{i} \rangle$ yields the thermal contribution to the stress \cite{wink:15}.

For passive systems with $v_0=0$, Eq.~(\ref{eq:virial}) reduces to the local virial expression  in thermal equilibrium presented in Ref.~\ocite{lion:12}. Compared to the \REV{global} virial \cite{wink:15}, additional terms with the derivative of $\Lambda_i$, $\dot \Lambda_i$, appear due to particles entering and leaving the local volume $\dV$. Most remarkably, activity yields a similar contribution as inertia, namely $\gamma \sum_i \langle v_{i \alpha}^a r_{i \alpha} \dot \Lambda_i \rangle / \gamma_R$, compared to $\sum_i \langle m \dot r_{i \alpha} r_{i \alpha} \dot \Lambda_i \rangle$. This demonstrates that transport  of ABPs across a plane yields a contribution to stress similarly to the momentum $\bm p_i^m = m \dot {\bm r}_i$ of a (passive) particle, in contrast to previous statements \cite{spec:16}. Hence, we can consider
\begin{align} \label{eq:moment}
\bm p_i^a = \frac{\gamma}{\gamma_R} \bm v_i^a
\end{align}
as an \REV{instantaneous} active momentum. \REV{Formally, an analogous momentum (impulse) has been introduced in Ref.~\ocite{fily:18} ``as momentum the active particle will receive on average from its active force in the future'', thus, the meaning is rather different.}

Equation~(\ref{eq:virial}) includes volume and boundary contributions. Thereby, terms with $\dot \Lambda_i$ are pure boundary terms. Following the procedure of Ref.~\ocite{lion:12}, the following external, $\sigma_{\alpha \alpha}^e$, and internal, $\sigma_{\alpha \alpha}^i$,  stress tensor components are identified,
\begin{align} \label{eq:stress_e}
\dV \sigma_{\alpha \alpha}^e = & \ \sum_{i=1}^N m \lla \dot{r}_i r_i \dot \Lambda_{i} \rra + \frac{\gamma}{\gamma_R} \sum_{i=1}^N   \lla  v^a_{i} r_{i} \dot \Lambda_{i} \rra
+ \sum_{i=1}^N \sum_{j=1}^N \lla (r_i- \lambda_{ij} r_{ij}) F_{ij} \Lambda_i (1-\Lambda_j)\rra
- \sum_{i=1}^N \sum_{j=1}^N \lla \lambda_{ij} r_{ij} F_{ij} (1- \Lambda_i) (1-\Lambda_j) \rra , \\
\label{eq:stress_i}
\dV \sigma_{\alpha \alpha}^i = & \ -  \sum_{i=1}^N m \lla \dot{r}_i^2 \Lambda_{i} \rra - \frac{\gamma}{\gamma_R} \sum_{i=1}^N  \lla v^a_{i} \dot r_{i} \Lambda_{i} \rra
 - \frac{1}{2} \sum_{i=1}^N \psum_{j=1}^N  \lla \lambda_{ij} r_{ij} F_{ij}   \rra .
\end{align}
Here, we assume that the volume $\dV$ is sufficiently far from the surface, and, hence, short-range wall forces do not contribute to the local stress in $\dV$. The external stress tensor  $\sigma_{\alpha \alpha}^e$ includes contributions by momentum flux across the boundary, with the active flux $\gamma \sum_{i=1}^N  \langle v^a_{i}  r_{i} \dot \Lambda_{i} \rangle/\gamma_R$, as well as boundary force terms; the third term on the right-hand side of Eq.~(\ref{eq:stress_e}) accounts for stress contributions by forces between particles $i$ inside  and $j$ outside of $\dV$, and the fourth term for force contributions, when all particles are outside of $\dV$, but the connecting line insects the volume $\dV$; $\lambda_{ij}$ denotes the fraction of the line connecting  particle $i$ and $j$ inside of the volume $\dV$  (cf. Fig.~\ref{fig:local_region}) \cite{lion:12}. Note, by suppression of the averages, instantaneous stresses are obtained.

The internal stress tensor $\sigma_{\alpha \alpha}^i$  (\ref{eq:stress_i}) includes kinetic contributions from inertia and activity of particles, which are inside the volume of interest, as well as interparticle forces. The interparticle-force term
\begin{align} \label{eq:stress_inter}
\dV \sigma_{\alpha \alpha}^{if} =
 - \frac{1}{2} \sum_{i=1}^N \psum_{j=1}^N  \lla \lambda_{ij} r_{ij} F_{ij}   \rra
\end{align}
includes contributions from particles both, inside of $\dV$, here $\lambda_{ij}=1$---this is the standard volume term---, as well as combinations, where a particle is inside and all  others, within the finite interaction range, are outside, or both are outside, but their connecting line intersects the volume $\dV$, here  $\lambda_{ij}<1$. The latter are boundary terms in analogy to the wall contributions in Eq.~(\ref{eq:stress_int}) of the \REV{global}  system with finite-range wall forces \cite{wink:93}.

Remarkably, a new expression for the swim stress is obtained
\begin{align} \label{eq:active_swim_stress}
\dV \sigma_{\alpha \alpha}^{s} = - \frac{\gamma}{\gamma_R} \sum_{i=1}^N \lla v^a_{i \alpha} \dot r_{i \alpha} \Lambda_{i \alpha} \rra ,
\end{align}
denote as local swim stress in the following,
different from the ``traditional'' swim stress of  Eq.~(\ref{eq:swim_stress_trad}).  Most importantly, the swim stress \eqref{eq:swim_stress_trad} vanishes in $\dV$,
and, hence, it is not the adequate expression for the calculation of the local active stress.
The vanishing   expression~\eqref{eq:swim_stress_trad} in a subvolume lead to the conclusion in Ref.~\ocite{spec:16} that the local pressure in an ideal active gas is identical with the passive-ideal-gas pressure when accounting for the respective local density. As a consequence, it was concluded that the active pressure is a boundary effect, because active forces were not contributing to momentum flux. Our calculations contradict this conclusions. First of all, we find an active flux, Eq.~(\ref{eq:moment}), and, secondly, another expression for the local active stress, Eq.~\eqref{eq:active_swim_stress}.  The vanishing average of Eq.~\eqref{eq:swim_stress_trad} in $\dV$ together with Eq.~\eqref{eq:virial_e} shows that the contribution by the active momentum is the negative of the active virial, i.e.,
\begin{align}
\sum_{i=1}^N  \lla p_{i \alpha}^a  r_{i \alpha} \dot \Lambda_{i \alpha} \rra = - \frac{\gamma}{\gamma_R} \sum_{i=1}^N \lla v^a_{i\alpha} \dot r_{i\alpha} \Lambda_{i \alpha} \rra .
\end{align}
However, if we extend the volume $\dV$ to the whole volume $V$, i.e., $\Lambda_i \equiv 1$, Eq.~(\ref{eq:orient}) in the Ito formulation yields $\lla v^a_{i \alpha} \dot r_{i \alpha} \rra = \gamma_R \lla v^a_{i \alpha} r_{i \alpha} \rra$, and Eqs.~\eqref{eq:stress_int} and \eqref{eq:stress_i} become identical. Note that the wall force has then to be taken into account in Eq.~\eqref{eq:stress_i}. Thus, our Eq.~ (\ref{eq:active_swim_stress}) is a  more general expression for the swim stress, since it applies locally as well as globally.

\subsection*{\REV{Global} stress of ideal ABP gas}

For an ideal  gas of ABPs, the interparticle forces are zero, i.e., $\bm F_{ij} = 0$, and Eq.~({\ref{eq:stress_int}) becomes
\begin{align}  \label{eq:stress_int_conf_gas}
V \sigma_{\alpha \alpha}^i =  - N k_B T-  \sum_{i=1}^N \gamma  \lla v^a_{i \alpha} r_{i \alpha} \rra  - \sum_{i=1}^N \lla F^w_{i \alpha}(r_{i \alpha}-S_{i \alpha}) \rra  ,
\end{align}
in the overdamped limit, $m=0$. According to Refs.~\ocite{wink:15,solo:15}, the swim stress (\ref{eq:swim_stress_trad}) can be written as
\begin{align}
\gamma \lla v^a_{i \alpha} r_{i \alpha} \rra  = \frac{\gamma}{d \gamma_R} v_0^2 + \frac{1}{\gamma_R} \lla F_{i \alpha}^w v^a_{i \alpha} \rra ,
\end{align}
where $d$ is the dimension, which leads to the expression
\begin{align}  \label{eq:stress_int_surf}
V \sigma_{\alpha \alpha}^i =  - N k_B T-  \frac{\gamma}{3 \gamma_R} N v_0^2 - \frac{1}{\gamma_R}\sum_{i=1}^N \lla F_{i \alpha}^w v^a_{i \alpha} \rra   -  \sum_{i=1}^N \lla F^w_{i \alpha}(r_{i \alpha}-L_{\alpha}) \rra
\end{align}
in 3D.
The first term on the right-hand side is the thermal contribution to stress, and the second term the ideal gas swim stress \cite{wink:15,solo:15,taka:14,mall:14,smal:15}
\begin{align} \label{eq:ideal_abp}
\sigma_{\alpha \alpha}^{id} = -  \frac{\gamma N}{3 \gamma_R V} v_0^2 .
\end{align}
Noteworthy, the wall interaction appears twice in Eq.~\eqref{eq:stress_int_surf}. The term proportional to the active velocity reflects and accounts for wall accumulation, whereas the other term captures finite-range effects of the wall force.

\subsection*{Local tress of ideal ABP gas}

For zero interparticle forces, Eq.~(\ref{eq:stress_i}) becomes
\begin{align} \label{eq:press_i}
\dV \sigma_{\alpha \alpha}^i =  - \sum_{i=1}^N m \lla \dot{r}_{i \alpha}^2 \Lambda_{i \alpha} \rra - \frac{\gamma}{\gamma_R}
\sum_{i=1}^N  \lla v^a_{i \alpha} \dot{r}_{i \alpha} \Lambda_{i \alpha} \rra .
\end{align}
With the formal solution of Eq.~(\ref{eq:abp_trans}),
\begin{align} \label{eq:int_velocity}
\dot r_{i \alpha} = \frac{1}{m} \int_{-\infty}^t e^{-\gamma(t-t')/m}   \left[\gamma v^a_{i \alpha}(t')  + F_{i \alpha}(t') + \Gamma_{i \alpha}(t') \right] dt',
\end{align}
we find
\begin{align}
\lla \dot{r}_{i \alpha}^2\rra =  \frac{k_BT}{m} + \frac{v_0^2}{d(1+m \gamma_R/\gamma)} , \hspace*{3mm} \lla v^a_{i \alpha} \dot{r}_{i \alpha}\rra =  \frac{v_0^2}{d(1+m \gamma_R/\gamma)}
\end{align}
in the stationary state.
In the evaluation of these expressions,  all wall contributions vanish as long as we assume that the volume $\dV$ is located far enough from a wall such that $e^{-\gamma(t-t^w_i)} \ll 1$, where $t_i^w$ is the time of the last encounter of an ABP with a wall before entering $\dV$. Thus, we obtain
\begin{align} \label{eq:stress_local_bulk}
\dV \sigma_{\alpha \alpha}^i = - N_{\dV} \left( k_BT  + \frac{\gamma}{d \gamma_R} v_0^2 \right) ,
\end{align}
with $N_{\dV}$ the number of ABPs in the volume $\dV$. The contributions with the Stokes number $m \gamma_R/\gamma$  cancel, and the result for the overdamped dynamics is obtained, as already discussed in Ref.~\ocite{taka:17}.  Evidently, the local stress is independent of the wall force. However, due to wall accumulation, the ABP density is not necessarily homogeneous over the whole volume, in particular, it is reduced in the bulk. Hence, walls can affect the bulk stress and pressure.


For an overdamped and athermal dynamics, Eq.~(\ref{eq:press_i})  becomes
\begin{align} \label{eq:press_i_lim}
\dV \sigma_{\alpha \alpha}^i =  - \frac{\gamma}{\gamma_R}
\sum_{i=1}^{N_{\dV}}  \lla (v^a_{i \alpha})^2 \Lambda_{i \alpha} \rra  = - N_{\dV} \frac{\gamma}{3 \gamma_R} v_0^2 \end{align}
in 3D by inserting  $\dot r_{i \alpha} = v^a_{i \alpha}$ from Eq.~(\ref{eq:abp_trans}),
Hence, stress is given by the square of the (active) velocity,  similar to the case of a  passive system.

\begin{figure}[t]
\centering
\includegraphics[width=0.5\columnwidth]{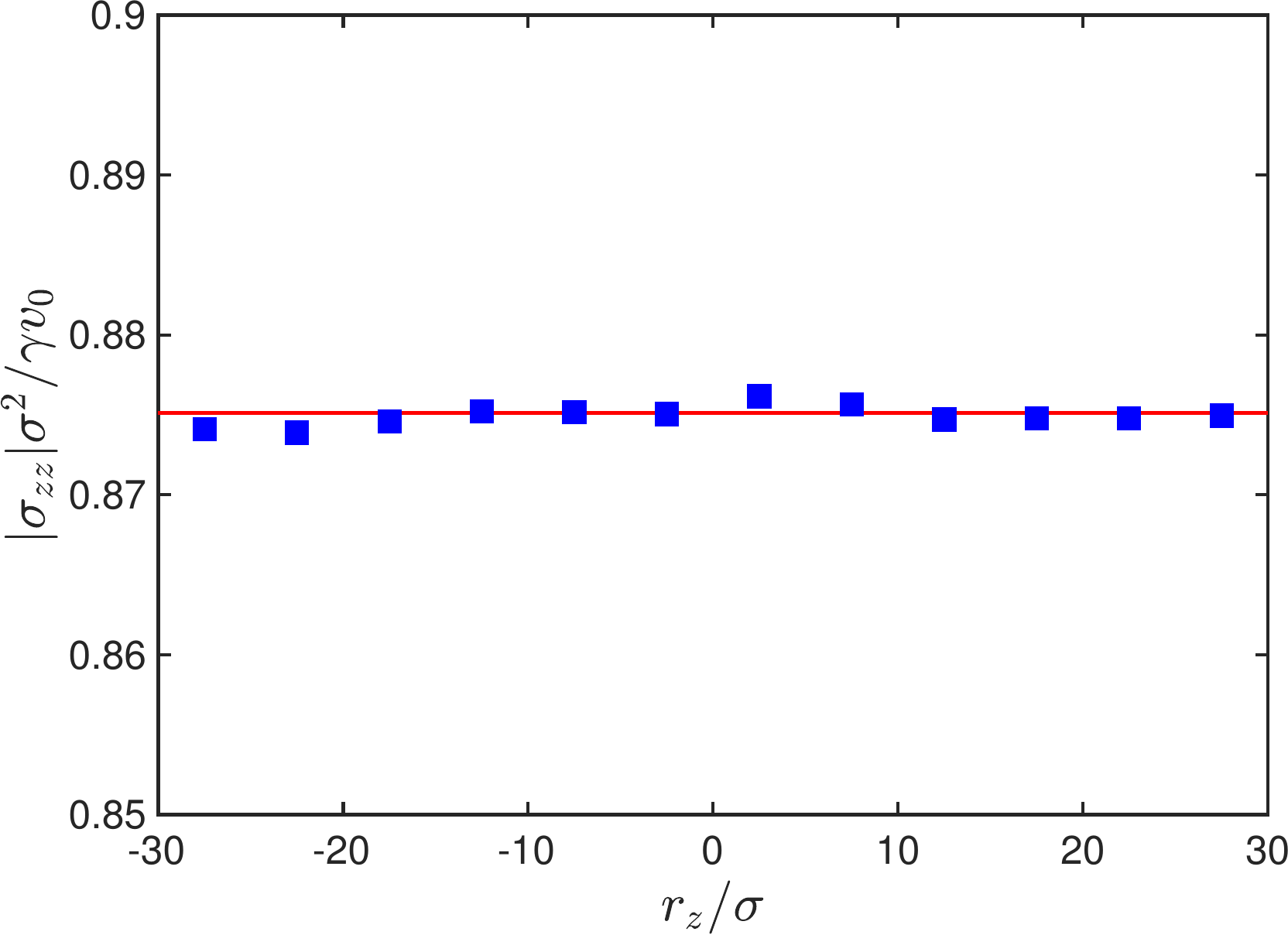}
\caption{Magnitude of the local stress $\sigma_{zz}$ \REV{according to Eq.~\eqref{eq:active_swim_stress}} in a system of non-interacting ABPs as a function of the center position $r_z$ of the local subvolume $\dV$ for $Pe=20$ (blue squares). The side length of the cubic volume $V$ is $L_\alpha = 100 \sigma$ and $\Delta L_z=0.05L_z$.  The solid line (red) indicates the virial stress of the volume $V$ with walls  (Eq.~(\ref{eq:stress_int_conf_gas}), $k_BT=0$).}
\label{fig:p_v_rz_wall}
\end{figure}

Evidently, Eqs.~(\ref{eq:stress_int_surf}) and \eqref{eq:stress_local_bulk} become identical in the limit  $L_\alpha \to \infty$, i.e.,  an infinite volume corresponding to the thermodynamic limit in case of a passive system. Then, the local particle density $N_{\dV}/\Delta V$ is equal to the density $N/V$, since finite-size effects and the inhomogeneities by surface accumulation become negligible compared to the overall particle number. However, the equality of the two expressions \eqref{eq:stress_int_surf} and \eqref{eq:press_i} applies for all surface separations as long as $(L_\alpha - \Delta L_\alpha)/l_p \gg 1$, where $l_p = v_0/\gamma_R$ is the persistence length of the active motion. The surface forces in Eq.~\eqref{eq:stress_int_surf} yield additional contributions to the ideal active gas stress. Associated bulk-density variations change the respective values of the local stress (\ref{eq:stress_local_bulk}).

\REV{Figure~\ref{fig:p_v_rz_wall} shows simulation results  for the local swim stress of non-interacting ABPs  in the subvolume $\dV$ according to Eq.~(\ref{eq:active_swim_stress}) as function of the center position $r_z$ of $\dV$, where  $\Delta L_x=\Delta L_y = 100\sigma$ and $\Delta L_z=0.05L_z$. Activity is characterized by the P\'eclet number $Pe=v_0/\sigma D_R$, where $Pe=20$ in Fig.~\ref{fig:p_v_rz_wall}.} In addition, the global wall stress according to  Eq.~(\ref{eq:stress_int_conf_gas}) ($k_BT =0$) is presented. Evidently, the local stress is independent of the location of the subvolume $\dV$ in the bulk of the system and it agrees with the wall stress within $1\%$.  Our results indicate that  stress in an active system is not a wall effect. According to Eq.~(\ref{eq:press_i_lim}), the local stress is proportional to the square of the active velocity of the ABPs within $\dV$, similar to the dependence on the kinetic energy in passive systems. Moreover, the simulations confirm the presence of an active momentum flux with the momentum of Eq.~(\ref{eq:moment}).

The dependence of the  stress on the activity is presented in Fig.~\ref{fig:p_v_pe_wall}(a). The local swim stress (\ref{eq:active_swim_stress}) coincides perfectly with the  wall stress (\ref{eq:stress_int_surf}) ($k_BT=0$) of the system. The scaled stress
\begin{align}
\frac{\sigma_{zz}^i \sigma^2}{\gamma v_0} = - \frac{\rho_{\dV} \sigma^3}{6} Pe ,
\end{align}
where $\rho_{\dV} = N_{\dV}/\dV$ is the bulk density,
decreases monotonically with increasing P\'eclet number, but deviates from the linear dependence of an unconfined (periodic boundary conditions) ideal ABP gas for large $Pe$.\cite{yan:15,ezhi:15,spec:16} The stress reduction is a consequence of an enhanced ABP surface accumulation with increasing $Pe$ and, correspondingly, a reduction of the bulk density. Here, we like to emphasize that the stresses calculated via Eqs.~(\ref{eq:stress_ext}) and (\ref{eq:stress_int}) perfectly agree with each other.

Figure~\ref{fig:p_v_pe_wall}(b) shows the relative density difference $\Delta \rho/\rho = (\rho -\rho_{\dV})/\rho$ with respect to the mean density $\rho = N/V$ as function of the P\'eclet number (squares). The simulation data, with the bulk density $\rho_{\dV}$, can well be described by the relation
\begin{align} \label{eq:density}
\frac{\Delta \rho}{\rho}  = \frac{1}{1+0.7 L_z/l_p} ,
\end{align}
which approaches zero in the limits $L_z\to \infty$ and $Pe \to 0$, and the bulk density becomes equal to the average density $\rho$ as for an passive ideal gas.
Such a dependence on $L_z/l_p$ has been obtained in Ref.~\ocite{spec:16} for the fraction of adsorbed ABPs at a wall. However, the numerical factor is somewhat different, we find $0.7$ compared to $0.6$ in  Ref.~\ocite{spec:16}. The difference could be related to the different adopted dimensionalities,  $d=3$ here versus $d=2$ in Ref.~\ocite{spec:16}. Moreover, Fig.~\ref{fig:p_v_pe_wall}(b) shows that the wall virial $\sum_i \lla  F_{i \alpha}^w v^a_{i \alpha} \rra / \gamma_R$ \eqref{eq:stress_int_surf} divided by the ideal ABP swim stress (\ref{eq:ideal_abp}) is identical with the relative density difference $\Delta \rho/\rho$ and, thus, exhibits the same dependence on the P\'eclet number (the finite-range wall-force contribution is negligible). Of course, this is expected by the equivalence of the stress expressions \eqref{eq:stress_int_conf_gas} and \eqref{eq:press_i}.

\begin{figure}[t]
\centering
\includegraphics[width=\columnwidth]{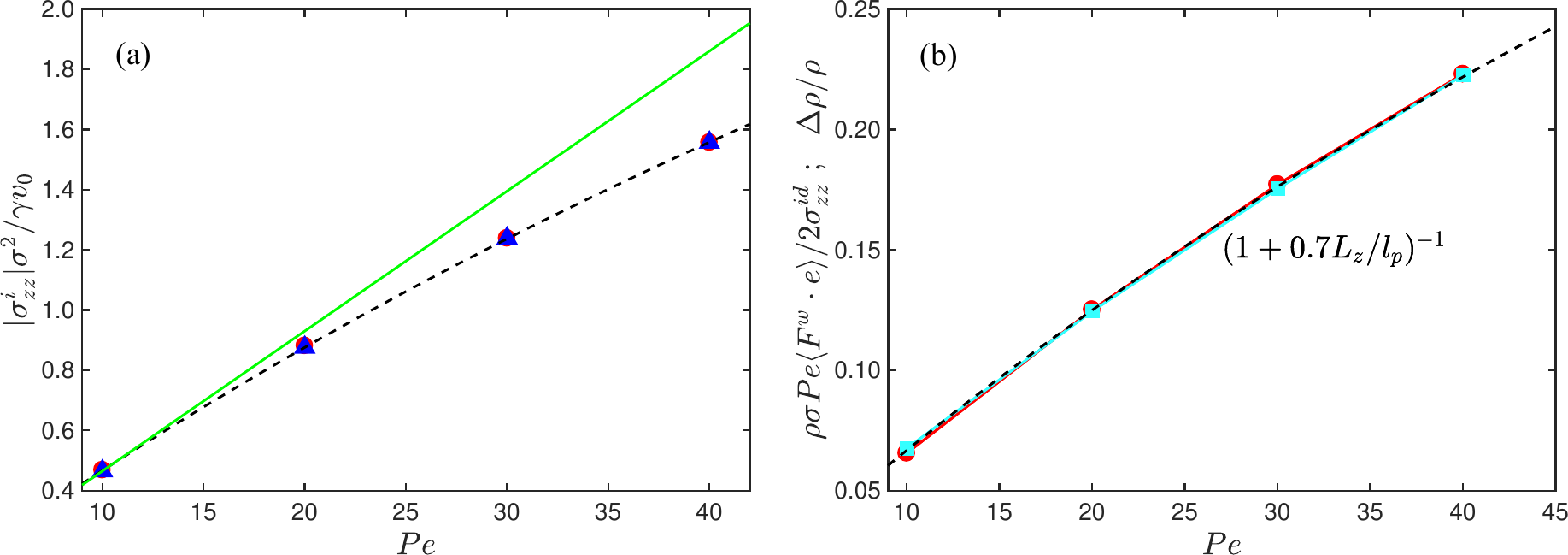}
\caption{(a) Magnitude of the local (blue triangles) and \REV{global} stress (red bullets)  in a system of non-interacting ABPs as a function of the P\'eclet number $Pe$. The solid line (green) indicates the ideal swim stress Eq.~\eqref{eq:ideal_abp}.
(b) Relative density $\Delta \rho/\rho$ (light blue squares) and the scaled virial of the wall force, $\sum_i \langle F_{i \alpha} v_{i \alpha }^a\rangle/\gamma_R$, as function of the P\'eclet number (red bullets). The dashed line represents the fit of Eq.~\eqref{eq:density}.
The wall separation is $L_z=100 \sigma$ and $\Delta L_z/L_z =  0.2$. Note that symbols overlap, emphasizing the agreement between the two ways of calculating stress. }
\label{fig:p_v_pe_wall}
\end{figure}

\begin{figure}[h]
\centering
\includegraphics[width=0.5\columnwidth]{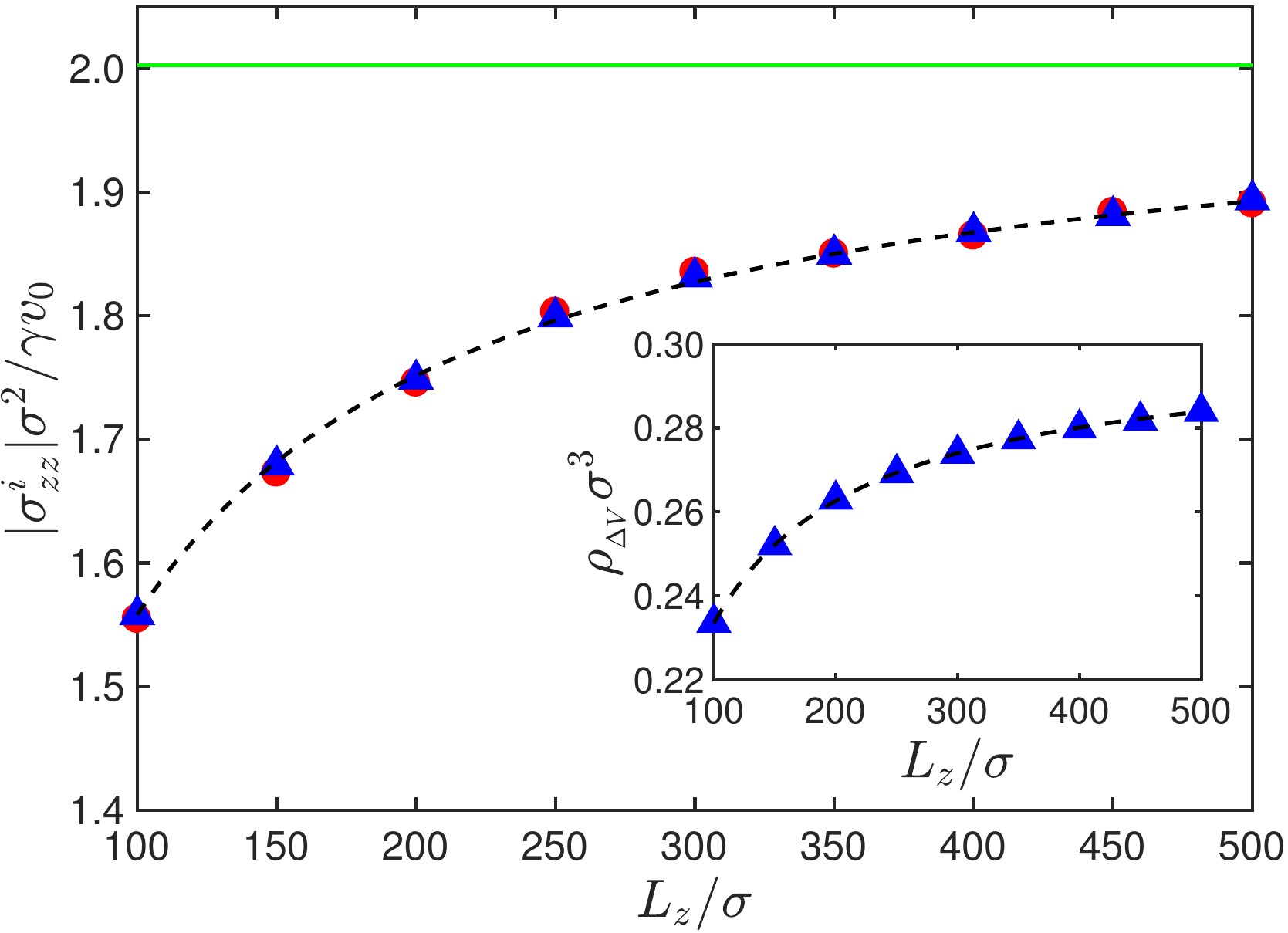}
\caption{Magnitude of the local  (blue triangles) and \REV{global} stress  (red bullets)  in a system of non-interacting ABPs as a function of the system size $L_z$.  The solid line (green) indicates the ideal swim stress. The dashed line indicates the stress $\sigma_{zz}^i=\sigma_{zz}^{id}\rho_{\dV}/\rho$, where $\Delta \rho/\rho$ is given by  Eq.~\eqref{eq:density}.  Inset: Variation of the bulk density with wall separation. The dashed line follows from Eq.~\eqref{eq:density}.
The P\'eclet number is $Pe = 40$ and $\Delta L_z/L_z =  0.2$.}
\label{fig:p_v_lz_wall}
\end{figure}

Figure~\ref{fig:p_v_lz_wall} illustrates the dependence of the stress on the wall separation $L_z$ for the P\'eclet number $Pe=40$. Again, simulation results for wall and local stress agree very well. With increasing system size, the wall effect decreases and the stress approaches the value of the ideal bulk stress. The relation between the local density and the system size is displayed in the inset of Fig.~\ref{fig:p_v_lz_wall}, which is well described by Eq.~\eqref{eq:density}.
The dependence of the ideal active gas stress on the P\'eclet number and system size is solely described by the respective dependence of the density $\rho_{\dV}$ on these quantities, since $\sigma_{zz}^i=\sigma_{zz}^{id} \rho_{\dV}/\rho$. This is confirmed in Figs.~\ref{fig:p_v_pe_wall}(a) and \ref{fig:p_v_lz_wall}.

Some of our findings seem to be rather evident as soon as the validity of the Eq.~\eqref{eq:press_i_lim} is taken for granted. However, this expression follows from our local stress with the local swim stress~\eqref{eq:active_swim_stress}, and not the stress expression for the system with walls. Specifically, Ref.~\ocite{spec:16} discusses wall-induced effects only. Hence, by our studies we emphasize the existence of a bulk stress and pressure and its identity with the respective wall values.

\begin{figure}[t]
\centering
\includegraphics[width=\columnwidth]{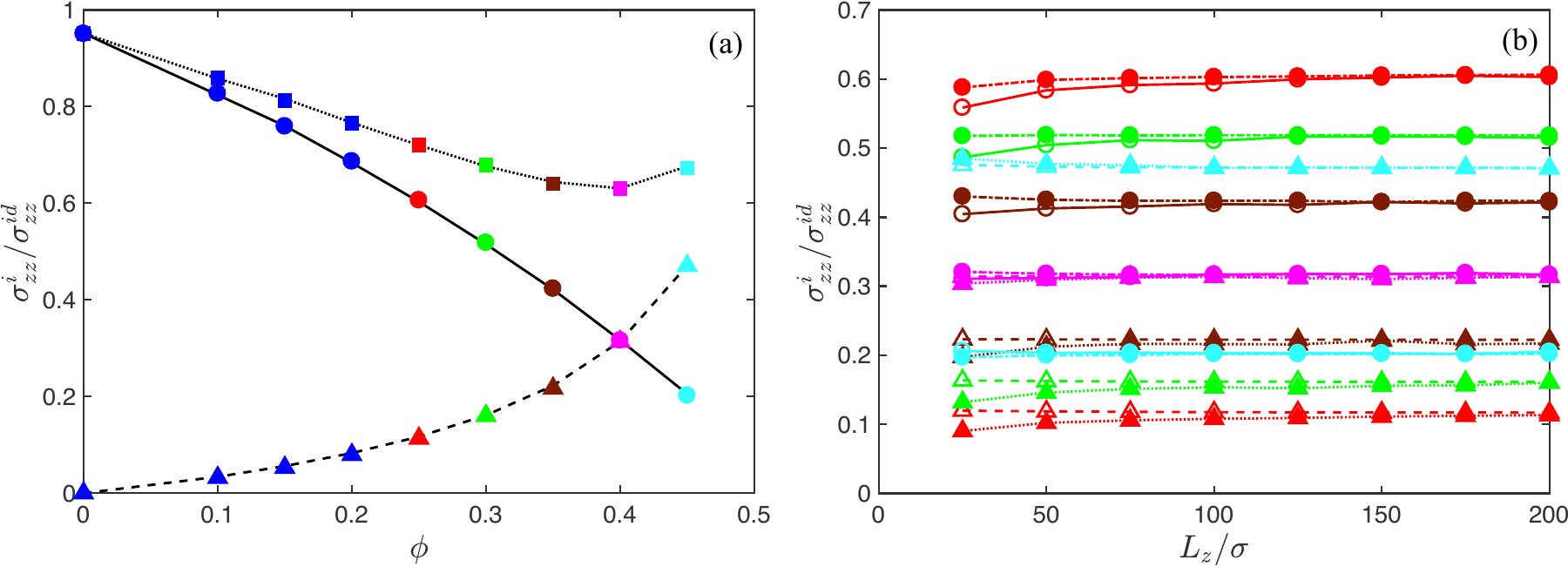}
\caption{(a) Relative local stress $\sigma_{zz}^i$ with respect to the active ideal gas stress $\sigma_{zz}^{id}$ in a system of  ABPs with excluded-volume interactions (dotted), the local swim stress \eqref{eq:active_swim_stress} (solid),  and the contribution by interparticle forces \eqref{eq:stress_inter} (dashed)  as a function of the ABP packing fraction $\phi$.    The colors refer to systems, where the system-size dependence of the various stress contributions is analyzed (see (b)). (b) Swim stress (bullets) and interparticle-force stress (triangles) calculated locally \eqref{eq:stress_i} (solid) and for the total system including walls \eqref{eq:stress_int} (open symbols) as a function of the wall separation $L_z$ for the concentrations $\phi = 0.25$, $0.3$, $0.35$, $0.4$, and $0.45$. The color code is the same as in (a). Note that the density-dependence changes nonmonotonically. The P\'eclet number is $Pe=10$, the wall separation $L_z=200\sigma$, and the width $\Delta L_z/L_z = 0.2$. The lines are guides for the eye.}
\label{fig:stress_ev_phi}
\end{figure}

\subsection*{Interacting ABPs}

Simulation results for the local interparticle force \eqref{eq:stress_inter}, local active \eqref{eq:active_swim_stress}, and global stress of ABP systems with excluded-volume interactions are depicted in Fig~\ref{fig:stress_ev_phi}(a) as function of the  packing fractions $\phi$. Again, we obtain excellent agreement between the various stress-tensor contributions calculated both, globally  and locally. Consistent with previous studies, the swim stress contribution decreases with increasing density, whereas the interparticle force contribution monotonically increases over the considered concentration range \cite{wink:15,marc:16.1,taka:16}. Above $\phi \approx 0.4$, the  contribution due to interparticle forces dominates the overall stress.
Figure~\ref{fig:stress_ev_phi}(b)  compares the local swim and interparticle stresses with those in the overall system for various wall distances and concentrations. Remarkably, the local swim stress (bullets) is independent of the wall separation, except for $\phi=0.25$. In this case, the volume $\dV$ is not sufficiently far away from the walls, and due to the comparably large persistence length $l_p=5\sigma$, the ABP velocity is affected by
wall interactions (cf. Eq.~\eqref{eq:int_velocity}). The effect vanishes with increasing concentration, since the persistence length decreases due to enhanced particle collisions at higher concentrations. The local swim stress in $\dV$ \eqref{eq:active_swim_stress} can be separated into two contributions by inserting the athermal and overdamped equation of motion  (Eq.~\eqref{eq:abp_trans}), namely
\begin{align} \label{eq:swim_stress_contr}
\sigma_{\alpha \alpha}^s = \sigma_{\alpha \alpha}^{id} - \frac{1}{\gamma_R} \sum_{i=1}^{N} \psum_{j=1}^N
F_{ij \alpha} v^a_{i \alpha} \Lambda_{i \alpha}.
 \end{align}
Hence, the reduction of the relative local swim stress in Fig.~\ref{fig:stress_ev_phi}(a) is a consequence of the correlations between the interparticle forces and the propulsion direction. The effect increases with increasing concentration.  We like to emphasize that density variations in the bulk are negligible small for the considered wall separations in Fig.~\ref{fig:stress_ev_phi}(b). In fact, surface accumulation is lower in a system of self-avoiding ABPs compared to an ideal ABP gas. The \REV{global} swim stress  \eqref{eq:swim_stress_trad} includes, in addition to the terms in Eq.~\eqref{eq:swim_stress_contr}, a wall contribution as shown in Ref.~\ocite{wink:15}. It is the latter part, which  implies a system-size dependence of the \REV{global} swim stress (open circles in Fig.~\ref{fig:stress_ev_phi}(b)). The effect gradually vanishes with increasing system size.
In contrast, the \REV{global} interparticle-force stress  is independent of system size, which can be expected for a system at constant average density $\rho$. However, the local interparticle-force stress depends on $L_z$. Since the total local and global stresses are equal, the system-size dependence of the local interparticle-force stress and that of the \REV{global} swim stress are equal.

Figure~\ref{fig:stress_ev_phi}(b) clearly reveals differences between the \REV{global} swim  (Eq.~\eqref{eq:stress_int}) and  local swim stress (Eq.~\eqref{eq:stress_i}), specifically for finite system sizes. However, in the thermodynamic  limit of infinitely large systems, the two definitions yield identical results. The total stress in a volume $V$ (global) or $\dV$ (local) is not affected by the difference, since other contributions by interparticle forces yield identical differences and, hence, compensate for the disparity in the swim  stresses.

\begin{figure}[t]
\centering
\includegraphics[width=\columnwidth]{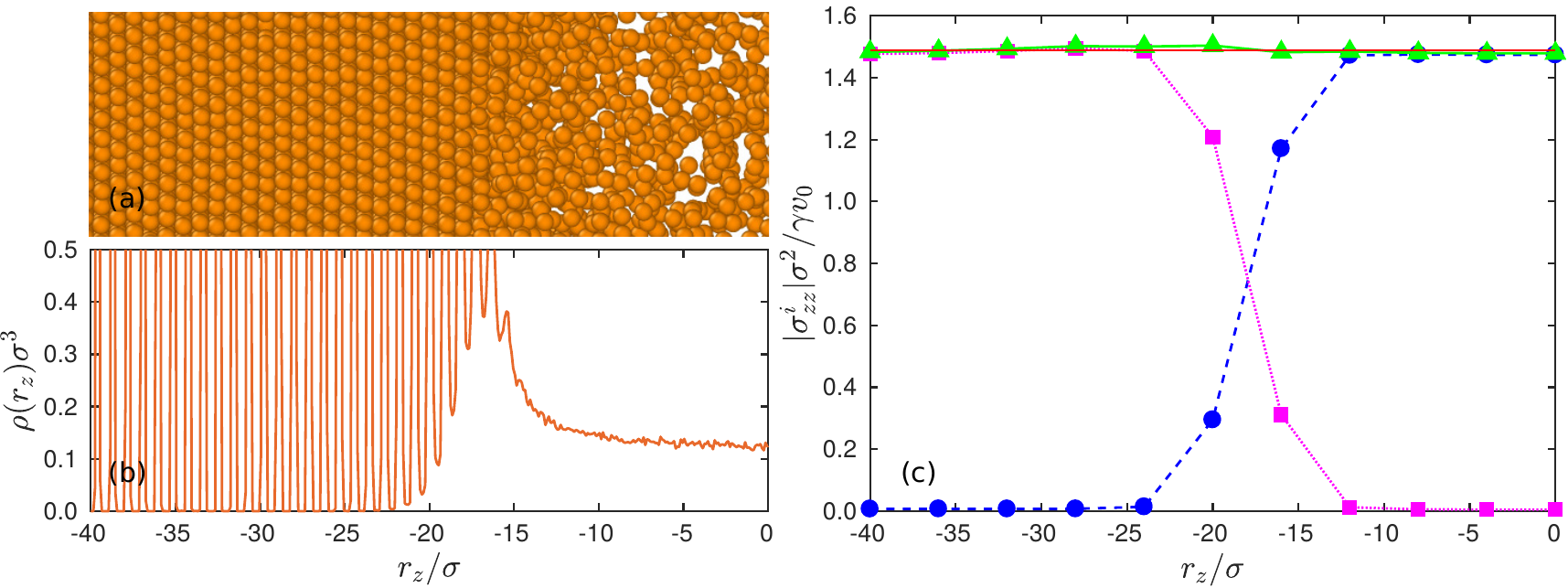}
\caption{Local \REV{(normal)} stress in an inhomogeneous ABP system confined between walls. (a) Snapshot of a section of the system illustrating the crystalline-like layers adjacent to a wall (left) and the fluid phase in the center (right).  (b) ABP-density distribution. Due to the pronounced layering adjacent to a wall, the density peaks are narrow and large, exceeding the $y$-scale by far.
(c) Contributions of the local swim stress \eqref{eq:active_swim_stress} (blue bullets) and the interparticle force stress \eqref{eq:stress_inter} (magenta squares) to the overall stress (green triangles). The solid red line indicates the  global stress  calculated via Eq.~(\ref{eq:stress_int}). The size of system is $L_x=L_y=25 \sigma$,  $L_z=100\sigma$, and the width of the local volume  $\Delta L_z=4 \sigma$.}
\label{fig:local_stress}
\end{figure}

\subsection*{Local stress in inhomogeneous systems of ABPs}

\REV{The stress in phase-separated, inhomogeneous ABP systems has been calculated before by various methods \cite{bial:15,solo:18}. The calculation of the global stress in Ref.~\ocite{bial:15} is performed with a particle-based model (ABPs) in a system with periodic boundary conditions. However, no stress normal to the interface has been obtained. In Ref.~\ocite{solo:18}, a continuum description based on a generalization of the Cahn-Hilliard equation is applied. Here, we illustrate the suitability of our particle-based expression for the calculation of local stresses  for a phase-separated system of ABPs confined between two walls. } The same geometry  is considered as before, but with the dimensions $L_x=L_y=25 \sigma$. The P\'eclet number $Pe= 80$ is sufficiently large to yield phase separation into crystalline-like layers adjacent to walls and a fluid phase in the center, as is displayed in Fig.~\ref{fig:local_stress}(a) and (b). In order to calculate the local stress, we consider local volumes of width $\Delta L_z= 4 \sigma$, i.e., we average over four crystalline layers. As shown in Fig.~\ref{fig:local_stress}(c), \REV{the stress normal to the walls and the interface ($z$ direction)} is constant over the whole system. Moreover, the local stress agrees with the stress exerted on the wall. In the various regimes, stress is dominated either by the swim stress (Eq.~\eqref{eq:active_swim_stress}) (central fluid part) or the force contribution to stress (Eq.~\eqref{eq:stress_inter}) (crystalline wall part). Specifically in the crossover regime, where both stresses contribute, the local stress is equal to the global stress (wall stress).
This result confirms that the derived expression is suitable for calculating local stress even in inhomogeneous systems.

\section*{Summary and conclusions} \label{sec:conclusions}

We have derived expressions for the stress (pressure) of a system of confined active Brownian particles, specifically, generally valid expressions for the local stress in a subvolume $\dV (\bm r)$ centered around a point $\bm r$. In addition to previously derived contributions to  the stress of active systems confined between walls \cite{wink:15}, the contribution of the finite-range surface interaction is taken into account in the calculation of the global stress. For the stress in a local volume ($\dV$), we find two expressions by following the strategy of Ref.~\ocite{lion:12}. On the one hand, stress is given by momentum transfer across a hypothetical plane, including force as well as kinetic (momentum) contributions. Here, we find a motility contribution by introducing an active momentum in analogy to the kinetic momentum of a particle of mass $m$.  On the other hand, we identify a virial expression for the active stress (local swim stress) of particles inside of $\dV$, different from the known swim stress \cite{taka:14,taka:14.1, wink:15}. \REV{In fact, in Ref.~\ocite{fily:18} a formally similar expression has been defined, however, with an active impulse of very different meaning.} Computer simulations show that our obtained local stress expression agrees quantitatively with the corresponding \REV{global} stress expression of an ABP system confined between  walls  for both, non-interacting  ABPs as well as ABPs interacting via a Lennard-Jones potential. This underlines the existence of an equation of state for a system of spherical ABPs.

Our results reach beyond previous considerations and discussions on the stress in active systems. Initially, swim pressure has been introduce via Clausius' virial \cite{beck:67}, taking active forces into account \cite{taka:14,yang:14.2}. Based on the equations of motion of ABPs for an infinite system, the swim pressure of an ideal active gas has then been proposed \cite{taka:14}. Subsequently, derivations of active pressure expressions have been presented for ABPs confined between walls \cite{wink:15,fala:16,spec:16} or in periodic systems \cite{wink:15,spec:16}. Alternatively, pressure expressions have been determined via the Fokker-Planck equation of an ABP system \cite{solo:15.1}, or the equations of motion for the density field \cite{fily:18}.  A derivation of an expression for a local pressure has be attempted in Ref.~\ocite{spec:16}, with the conclusion that a local pressures exists, but the bulk value differs from the mechanical pressure exerted on confining walls. In case of an ideal gas of ABPs, the bulk pressure would be given by $\rho(\bm r) k_BT$, with the position dependent density $\rho (\bm r)$ and, hence, would vanish in an athermal system. Hence, active pressure is considered as a boundary effect \cite{spec:16}.
In contrast, the present calculations provide a general expression for the local stress and pressure and thus, mark a substantial advancement. The derived local stress/pressure expressions are more general and reduce to the known expressions in the limit of the local volume being equal to the overall volume. This is emphasized by the
fact that the local swim stress \eqref{eq:active_swim_stress} is finite, while the ``traditional'' swim stress \eqref{eq:swim_stress_trad} vanishes in $\dV$.
Most importantly, our calculations demonstrated that active stress/pressure  is not a wall (boundary) effect, as stated in Ref.~\ocite{spec:16}, but is caused by active momentum transport in analogy to  momentum transfer in a passive system. Hence, our results allow us to resolve apparent contradictions in the interpretation of local stress and consequently in the existence of equations of state in active systems.

As we demonstrated, local stress or pressure can be calculated in inhomogeneous ABP systems. Extensions to local shear stresses are straightforward and, thus, surface tensions of active systems can be determined.

\section*{Methods}\label{sec:simulation_method}}

We perform simulations in three dimensions considering the overdamped Eq.~(\ref{eq:abp_trans}),  $m=0$, without thermal fluctuations. The translational equations of motion~(\ref{eq:abp_trans}) are solved via the Ermak-McCammon algorithm~\cite{erma:78}. The equations of motion (\ref{eq:orient}) for the orientation vectors are solved using the scheme described in Ref.~\ocite{wink:15}.

Pair-wise interparticle interactions are taken into account by the repulsive Lennard-Jones potential
\begin{align} \label{eq:lj_pot}
U_{LJ} = \left\{
\begin{array}{cc} \epsilon \left[ \displaystyle \left(\frac{\sigma}{r_{ij}}\right)^{12} -  \left(\frac{\sigma}{r_{ij}}\right)^{6} + \frac{1}{4}\right] , & r_{ij} \le \sqrt[6]{2} \sigma \\ 0, &  r_{ij}>\sqrt[6]{2} \sigma
\end{array}
\right.,
\end{align}
where $r_{ij} = |\bm r_{ij}|$ and the interaction strength is set to $\epsilon/k_BT = Pe$; $Pe$ is P\'eclet number $Pe$ defined as
\begin{align} \label{eq:peclet}
Pe  = \frac{ v_0}{ \sigma D_R} .
\end{align}
This yields the same particle overlap independent of the P\'eclet number \cite{fily:14.1}.
The ABPs are confined between two  walls parallel to the $xy$-plane of the Cartesian reference frame located at $S_{iz}=\pm L_z/2$ (cf. Fig.~\ref{fig:local_region}). Parallel to the walls, periodic boundary conditions are applied. The purely repulsive ABP-wall interaction is described by the Lennard-Jones potential of Eq.~\eqref{eq:lj_pot} with $r_{ij}$ replaced by $z_i-S_{i z}$.

We determine the stress $\sigma_{zz}$ along the $z$-axis in two ways. Firstly, we calculate the \REV{global} stress of the whole system according to Eq.~(\ref{eq:stress_int}) ($m=0$). Secondly, the local stress in a volume $\Delta V$ is calculated via Eq.~(\ref{eq:stress_i}).
Thereby,  the box dimension $L_z$ normal to the walls is chosen such that it is significantly larger than the persistence length $l_p=v_0/2D_R=Pe\sigma/2$ of an ABP, in particular, we consider wall separations $L_z \geqslant 5l_p$.
The volume $\Delta V$ of width $\Delta L_z$ is located in the bulk of the system along the $z$-axis (cf. Fig.~\ref{fig:local_region}). If not indicated otherwise, the dimensions of the volume $V$ are $L_x=L_y = 40 \sigma$, $L_z=100 \sigma$, and $\Delta L_z=0.2 L_z$. In Fig.~\ref{fig:local_region},  $r_z$ denotes the position of the center of the slit within the interval $L_z/2 < r_z < L_z/2$. The number  density of ABPs in $V$ is $\rho\sigma^3=0.3$.

\section*{Acknowledgements}

Financial support by the Deutsche Forschungsgemeinschaft (DFG) within the priority program SPP 1726 ``Microswimmers—from Single Particle Motion to Collective Behaviour'' is gratefully acknowledged.





\end{document}